\documentclass[a4paper]{jpconf}
\usepackage{graphicx}
\usepackage{mathrsfs}
\usepackage{amssymb}
\usepackage{amsmath}
\usepackage{graphicx}
\usepackage{bm}

\begin{document}
\title{Directional detection of Dark Matter with MIMAC: WIMP identification and track reconstruction}

\author{J.~Billard, F.~Mayet, C.~Grignon and D.~Santos}

\address{Laboratoire de Physique Subatomique et de Cosmologie, Universit\'e Joseph Fourier Grenoble 1,
  CNRS/IN2P3, Institut Polytechnique de Grenoble, Grenoble, France}

\ead{billard@lpsc.in2p3.fr}

\begin{abstract}
Directional detection is a promising Dark Matter search strategy. Indeed, WIMP-induced recoils present a direction dependence toward the Cygnus constellation, while 
background-induced recoils exhibit an isotropic distribution in the galactic rest frame. 
Taking advantage on these characteristic features and even in the presence of a sizeable background, we show for the first time the possibility to 
constrain the WIMP properties, both from particle and galactic halo  physics, 
leading to an identification of non-baryonic Dark Matter. However, such results need highly accurate track reconstruction which should be reachable by the MIMAC detector
using a dedicated readout combined with a likelihood analysis of recoiling nuclei.
\end{abstract}

\section{Introduction}

Taking advantage of the astrophysical framework, directional detection of Dark Matter is an interesting strategy in order to distinguish
 WIMP events from background ones \cite{spergel}.
Indeed, like most spiral galaxies, the Milky Way is supposed to be immersed in a halo of WIMPs which outweighs the luminous component by at 
least one order of magnitude. As the Solar System rotates around the galactic center through this Dark Matter halo, WIMPs should mainly come
 from the direction to which points the
Sun velocity vector and which happens to be roughly in the direction of the Cygnus constellation ($\ell_\odot = 90^\circ,  b_\odot =  0^\circ$).
  Hence, we argue that a clear and unambigous signature of a Dark Matter
  detection could be done by showing the correlation of the measured signal with the direction of the solar motion.\\
 
\begin{figure}[t]
\begin{center}
\includegraphics[scale=0.4]{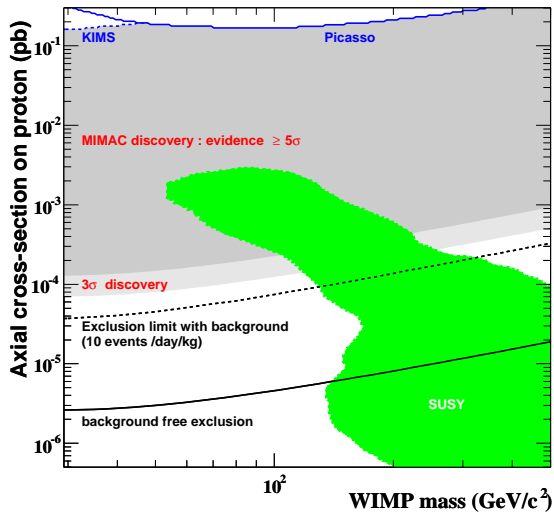}
\caption{Spin dependent cross-section on proton (pb) as a function of the WIMP mass ($\rm GeV/c^2$). 
Exclusion limits from some direct detection experiments are presented, KIMS~\cite{kims} and Picasso~\cite{picasso} as well as
the theoretical region, obtained within the framework of the constrained minimal supersymmetric model from \cite{superbayes}.
Contours corresponding to a significance greater than 3$\sigma$ and 5$\sigma$ are presented in dark and light grey.
The exclusion limit corresponding to pure background data is presented as the black dashed line and the detector sensitivity as the black solid line.}
\label{fig:discovery}
\end{center}
\end{figure}

 Several project of directional detectors are being developed \cite{white,santos} and some of them are already taking data \cite{DMTPC,DRIFT,NEWAGE}.
  The first step when analysing directional data should be to look for a signal pointing toward the
  Cygnus Constellation with a sufficiently high significance \cite{billard.disco}. If no evidence in favor of a Galactic origin of the signal is deduced from the previous
 analysis, then an exclusion limit should be derived \cite{billard.exclusion}.
On figure \ref{fig:discovery} we present the projected discovery regions and exclusion limits for a forthcoming directional detector proposed by the MIMAC
 collaboration. We consider 
a  10 kg $\rm CF_4$ detector 
operated during  $\sim 3$ years,  allowing 3D 
track reconstruction, with a $10^\circ$ angular resolution, a  recoil energy range 5-50 keV and with a conservative background rate of 10 evts/kg/year.
From figure \ref{fig:discovery}, two different scenarii may be distinguished:
\begin{itemize}
\item The dark and light-grey shaded areas represent the contours where a directional
detection of Dark Matter would have a significance greater than 3$\sigma$ and 5$\sigma$. These contours are deduced from the map-based likelihood method \cite{billard.disco}.
As an illustration, if the WIMP-nucleon cross-section is about $10^{-4}$ pb with a WIMP mass of 100 GeV.c$^{-2}$, the detector would have a
 Dark Matter detection with a significance greater than 3$\sigma$. 
\item If the WIMP-nucleon cross-section is lower than $10^{-5}$ pb, then an exclusion limit is deduced using the extended likelihood method (black dashed line) presented in
 \cite{billard.exclusion}. As a benchmark
and to illustrate the effect of background on exclusion limits, the detector sensitivity (no event) is presented (black solid line).
\end{itemize} 
Figure \ref{fig:discovery} also presents exclusion limits from direct detection experiments, KIMS~\cite{kims} and Picasso~\cite{picasso} as well as
the theoretical region, obtained within the framework of the constrained minimal supersymmetric model taken from \cite{superbayes}. We can conclude that a directional
detector like MIMAC will cover an important region of interest worth being investigated.\\

Constraining Dark Matter parameters both from particle physics (mass $m_\chi$ and cross section $\sigma_n$) and galactic halo (WIMP velocity distribution)
 with upcoming Dark Matter experiments is a main concern of 
current phenomenological studies. In the context of
   upcoming experiments which might  give  WIMP positive detection ($\sigma_n \sim 10^{-3}$ pb) instead of background rejection, it is of  
  particular interest   to develop a model-independent formalism to constrain Dark Matter properties.
  A new approach has been presented in \cite{billard.MCMC}. The idea is to constrain Dark Matter properties using a high dimensional multivariate analysis within
   the framework of a general halo model. This strategy is referred to as {\it quasi model-independent} method 
   as all parameters are constrained directly from the data of a single experiment.\\
   Indeed, directional detection presents a high identification potential thanks to  the combined measurement of the recoil energy $E_R$ and 
   the recoil direction ($\ell_R,b_R$). It allows to achieve multivariate analysis in order to obviate systematic bias  
   in the determination of the WIMP properties, due to wrong halo model assumptions~\cite{green.masse1}.\\

\section{Identification of Dark Matter}

 Within the framework of a quasi model-independent
method and by using a Markov Chain Monte Carlo analysis of recoil events,  we show for the first time the possibility to 
contrain the  WIMP properties, 
both from particle physics ($m_\chi, \sigma_n$) and galactic Dark Matter halo physics (velocity dispersions). 
This leads to an identification of non-baryonic Dark Matter, which could be reached within few years by upcoming directional detectors \cite{white}.

\subsection{Directional framework}

Directional detection depends   crucially on the WIMP velocity distribution. The isothermal sphere halo model is 
often considered but it is worth going beyond this standard paradigm in the case of a model-independent analysis. 
Indeed, recent  results from 
N-body simulations are in favor of triaxial Dark Matter halos with anisotropic velocity distributions~\cite{nezri}. Moreover, recent observations of Sagittarius stellar
tidal stream have shown evidence for a triaxial Milky Way Dark Matter halo \cite{Law:2009yq}.

The multivariate Gaussian WIMP velocity distribution  \cite{Evans:2000gr}  corresponds  to the generalization of the standard isothermal sphere with a density profile 
$\rho(r)\propto 1/r^2$, leading to a smooth WIMP velocity distribution,  
a flat rotation curve and no substructure. The WIMP velocity distribution in 
the laboratory frame is   given by,
\begin{equation}
f(\vec{v}) = \frac{1}{(8\pi^3\det{\boldsymbol\sigma}^2_v)^{1/2}}\exp{\left[-\frac{1}{2}(\vec{v} - \vec{v}_{\odot})^T {\boldsymbol\sigma}^{-2}_v(\vec{v} - \vec{v}_{\odot})\right]}
\end{equation}
where ${\boldsymbol\sigma}_v = \text{diag}[\sigma_{x}, \sigma_{y}, \sigma_{z}]$ is the velocity dispersion tensor 
assumed to be diagonal in the Galactic rest frame ($\hat{x}$, $\hat{y}$, $\hat{z}$) and $\vec{v}_{\odot}$ is the Sun motion with respect to
the Galactic rest frame. When neglecting the Sun peculiar velocity and the Earth orbital 
velocity about the Sun,  $\vec{v}_{\odot}$ corresponds to the detector velocity in
the Galactic rest frame and is taken to be $v_{\odot} = 220$ km.s$^{-1}$ along the $\hat{y}$ axis pointing toward the Cygnus constellation at 
($\ell_{\odot} = 90^{\circ}$, $b_{\odot} = 0^{\circ}$). 
 The velocity anisotropy $\beta(r)$, is then defined \cite{biney} as
  \begin{equation}
  \beta(r) = 1 - \frac{\sigma^2_{y} + \sigma^2_{z}}{2\sigma^2_x}
  \label{eq:beta}
  \end{equation}
According to N-body simulations, the $\beta$ parameter at the Solar radius   spans the range $0-0.4$, corresponding to 
radial anistropy.\\
In the following, 
the input halo model  used to generate simulated data, is chosen according to two models : a standard isotropic halo ($\beta=0$) in which case the velocity dispersions are
related to the local circular velocity $v_0 = 220$ km/s as $\sigma_{x} = \sigma_{y} = \sigma_{z} = v_0/\sqrt{2}$;
and an extremely anisotropic halo ($\beta = 0.4$), 
with the following velocity dispersions $\{\sigma_x = 200$ km/s; $\sigma_z = 169$ km/s; $\sigma_y = 140$ km/s$\}$. 
The latter case corresponds to  the logarithmic ellipsoidal halo model from \cite{Evans:2000gr} with the 
Sun located on the major axis of the halo with the axis ratios $p$ and $q$ equal to 0.9
and 0.8 respectively.\\  
  
The  directional recoil  rate  is given by  \cite{gondolo} :
\begin{equation}
\frac{\mathrm{d}^2R}{\mathrm{d}E_R\mathrm{d}\Omega_R} = \frac{\rho_0\sigma_0}{4\pi m_{\chi}m^2_r}F^2(E_R)\hat{f}(v_{\text{min}},\hat{q}),
\label{directionalrate}
\end{equation}
with $m_{\chi}$ the WIMP mass, $m_r$ the WIMP-nucleus reduced mass, $\rho_0=0.3 \ {\rm GeV/c^2/cm^3}$   the local Dark Matter density, $\sigma_0$   the
WIMP-nucleus elastic scattering cross section, $F(E_R)$  the form factor  (using the axial expression from \cite{lewin}) and  
$v_{\text{min}}$ the   minimal WIMP velocity required to produce a
nuclear recoil of energy $E_R$. 
Finally, $\hat{f}(v_{\text{min}},\hat{q})$ is the three-dimensional Radon transform of the WIMP 
velocity distribution $f(\vec{v})$.  Using the Fourier slice theorem \cite{gondolo},    the Radon transform of the 
multivariate 
Gaussian is,
\begin{equation}
\hat{f}(v_{\text{min}},\hat{q}) = \frac{1}{(2\pi\hat{q}^T{\boldsymbol\sigma}^2_v\hat{q})^{1/2}}\exp{\left[-\frac{\left[v_{\text{min}} - \hat{q}.\vec{v}_{\odot}\right]^2}{2\hat{q}^T{\boldsymbol\sigma}^2_v\hat{q}}\right]}.
\end{equation}

In the following, the model is characterized by 8 free parameters which are 
$\{m_{\chi}, \log_{10}(\sigma_n), l_{\odot}, b_{\odot},\sigma_{x}, \sigma_{y}, \sigma_{z}, R_b\}$, 
where the direction $(l_{\odot}, b_{\odot})$ refers to 
the main direction of the recorded events \cite{billard.disco}, $\sigma_n$ is the WIMP-nucleon cross section directly related to $\sigma_0$ in the 
pure proton approximation and $R_b$ is the background rate.
We have considered flat prior for each parameter. In such case, the Bayes'
 theorem is simplified and the target distribution   reduces to the   8 dimensional 
 likelihood function $\mathscr{L}(\vec{\theta})$ dedicated to unbinned data as,
  \begin{equation}
 \mathscr{L}(\vec{\theta}) = \frac{(\mu_s + \mu_b)^N}{N!}e^{-(\mu_s + \mu_b)} \ \times \ \prod_{n = 1}^{N_{\text{event}}} \left[ \frac{\mu_s }{\mu_s + \mu_b} S(\vec{R}_n)  + \frac{\mu_b }{\mu_s + \mu_b}B(\vec{R}_n)\right ]
 \end{equation}
 where $\mu_s$ and $\mu_b = R_b\times\xi$ are the expected number of WIMP   and background events respectively. $\vec{R}_n$ refers to the  
 energy and direction of each event while the functions $S$ and $B$ are the directional   rate of the WIMP   
 and the background events respectively. This MCMC analysis is based on the Metropolis-Hastings algorithm, using chain 
 subsampling   according to  the burn-in and  correlation lengths to deal only with 
 independent samples \cite{billard.MCMC}.

\subsection{Results from a benchmark model}

We first applied this MCMC method on a benchmark model with the following characteristics~: 
isotropic halo model ($\beta=0$) with a sizeable background contamination ($10 \ {\rm kg^{-1}.year^{-1}}$), a  $50 \ {\rm GeV/c^2}$ WIMP and
 a  WIMP-nucleon axial cross section 
$\sigma_n = 10^{-3}$ pb, well below current exclusion limits, see figure \ref{fig:discovery}. The input model is used to generate simulated 
data considering a MIMAC like detector as described above. These data are then analysed with the directional MCMC method.
 The full MCMC result with 2D correlations may be found in \cite{billard.MCMC}. However, major results are shown on table~\ref{tab:modelinputDark} and the conclusion 
is threefold:
\begin{itemize}
\item the discovery proof is given by the reconstruction of the main incoming direction $(\ell_\odot,b_\odot)$ pointing
toward the Cygnus constellation within 2.5$^{\circ}$, which is in favor of a positive Dark Matter detection.
\item The three velocity dispersions are strongly and consistently constrained according the input values corresponding to an isothermal sphere. Moreover, using 
equation~(\ref{eq:beta}), we can evaluate the posterior Probability Density Function (PDF) of the velocity anisotropy parameter ($\beta$) which is in favor of an isotropic
halo ($\beta = -0.073^{+0.29}_{-0.18}$).
\item As the galactic Dark Matter halo properties are well constrained, constraints on the WIMP properties ($m_{\chi}, \sigma_n$) are also strong and
consitent according to the input WIMP parameters: ($m_{\chi} = 51.8^{+5.6}_{-19.4}$ GeV/c$^2$, $\log_{10}(\sigma_n) = -3.01^{+0.05}_{-0.08}$ pb).
\end{itemize} 

As a conclusion of this benchmark study, we can deduce that this MCMC analysis, combined with directional data,
 is very well suited to constrain Dark Matter properties in a quasi model-independent way.

\setlength{\tabcolsep}{0.1cm}
\renewcommand{\arraystretch}{1.4}
\begin{table}
\caption{Comparison of the values of the parameters for the 
input model (isotropic halo and a flat background) and as extracted after the MCMC analysis from the marginalized distributions. 
We quote mean value of the PDF distribution and (68 \% CL) error bars.}
\label{tab:modelinputDark}
\begin{center}
{\scriptsize
\hspace*{-1cm}
\begin{tabular}{llllllllll}
\br
& $m_{\chi} \ {\rm (GeV/c^2)}$ &  $\log_{10}(\sigma_n) \ {\rm (pb)}$ & $ \ell_{\odot} \ {\rm (^\circ)}$ & $b_{\odot} \ {\rm (^\circ)}$ 
 & $\sigma_{x} \ {\rm (km.s^{-1})}$ & $\sigma_{y} \ {\rm (km.s^{-1})}$ & $\sigma_{z} \ {\rm (km.s^{-1})}$ & $\beta$ & $R_b \ 
 {\rm (kg^{-1}year^{-1})}$ \\
\mr
Input & 50    &   -3  & 90 & 0 & 150 & 150 & 150 & 0 & 10 \\
Output &  $51.8^{+5.6}_{-19.4}$    & $-3.01^{+0.05}_{-0.08}$  &  $92.2^{+2.5}_{-2.5}$ & $2.0^{+2.5}_{-2.5}$ & $158^{+15}_{-17}$ & $164^{+27}_{-26}$ & $145^{+14}_{-17}$ & $-0.073^{+0.29}_{-0.18}$ & 
$10.97 \pm 1.2 $ \\
\br
\end{tabular}
}
\end{center}
\end{table}
\renewcommand{\arraystretch}{1.1}

\subsection{Varying the input parameters}

\begin{figure*}[t]
\begin{center}

\includegraphics[scale=0.39,angle=0]{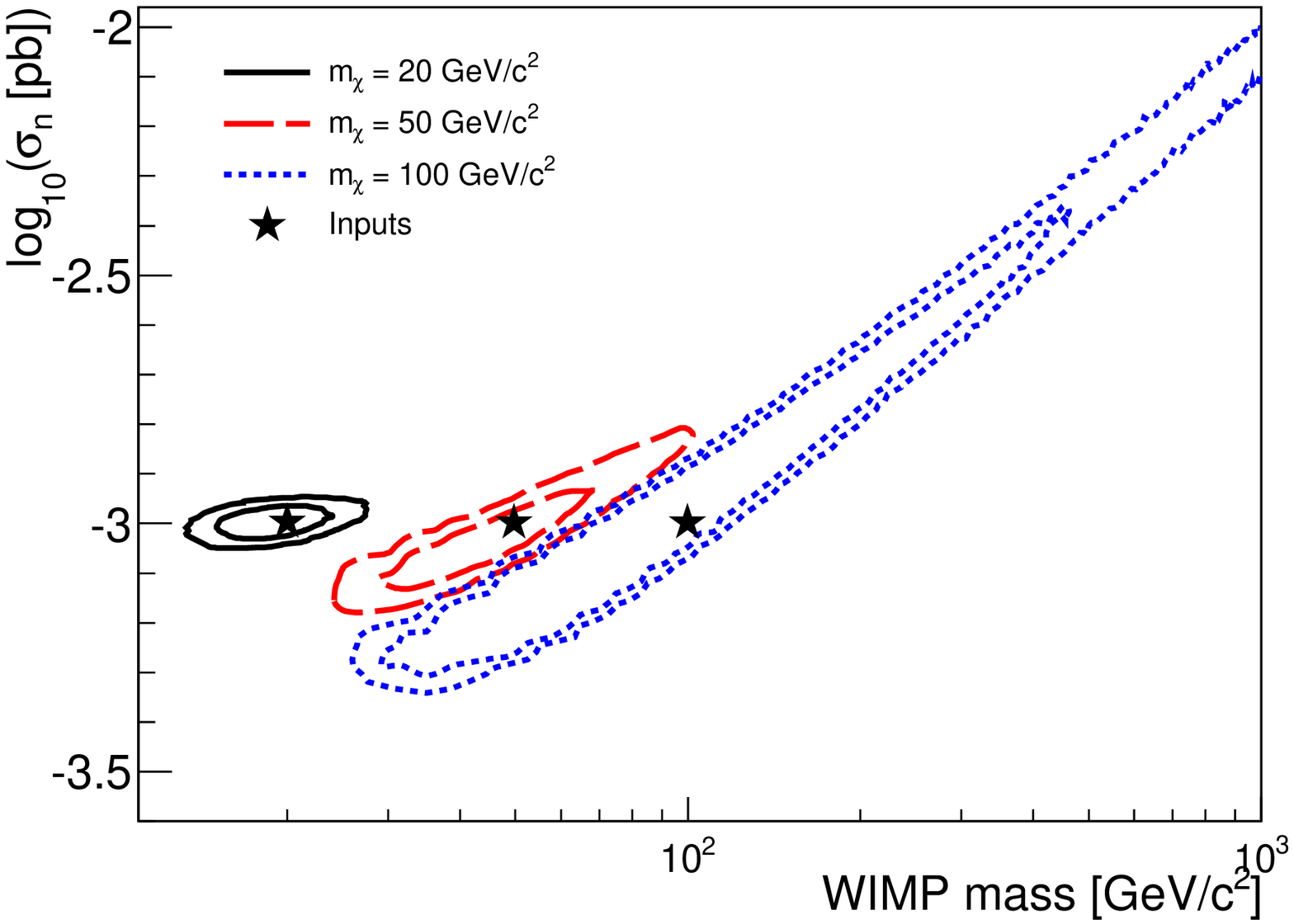}
\includegraphics[scale=0.39,angle=0]{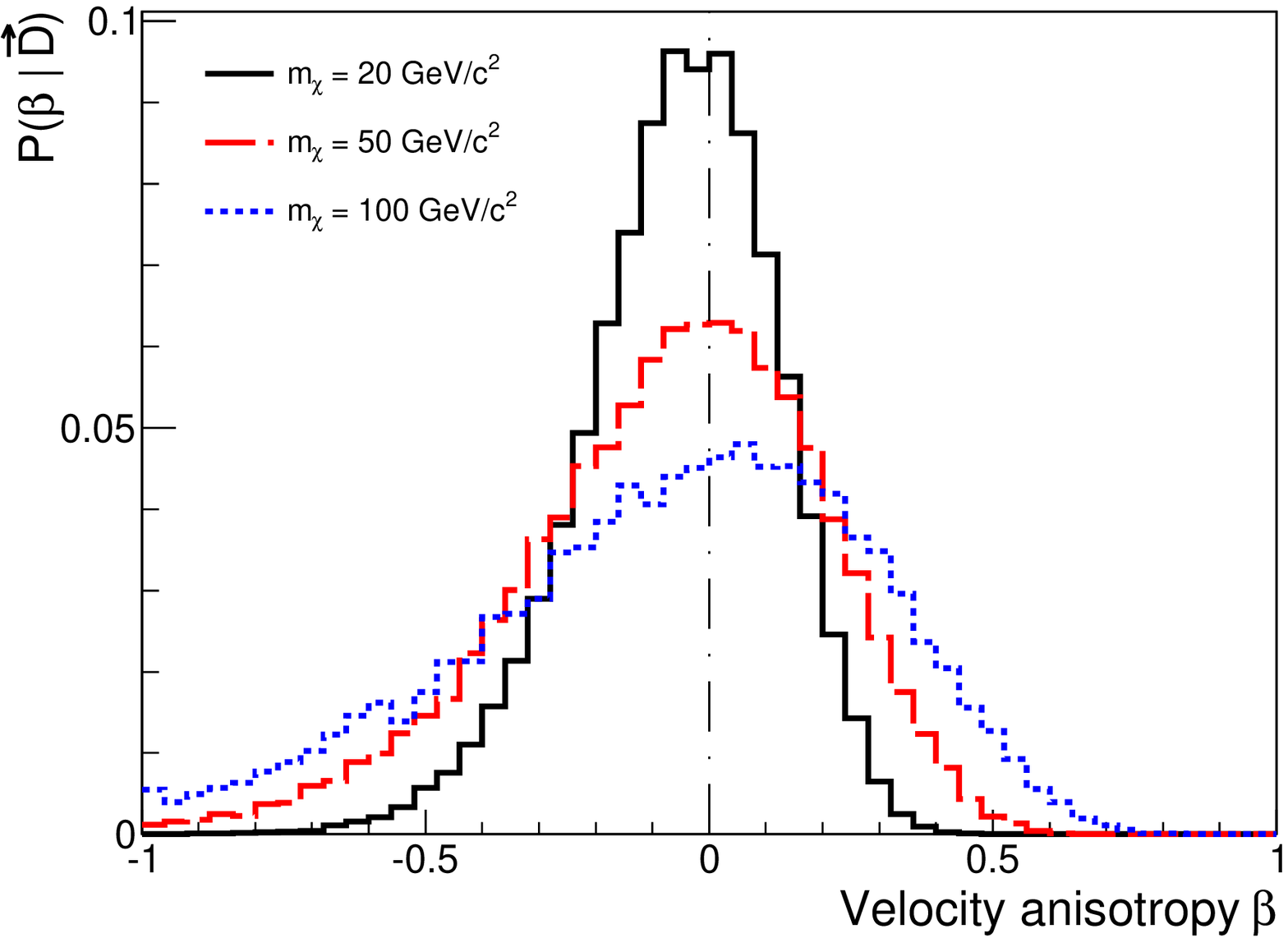}
\caption{Left panel : 68\% and 95\% contour level in the ($m_{\chi},\sigma_n$) plan, for the isotropic input model and for a WIMP mass 
equal to 20, 50 and 100 $\rm GeV/c^2$. 
Right panel : posterior PDF distribution of the $\beta$ parameter for the same models.} 
\label{fig:WIMPMass}

\includegraphics[scale=0.39,angle=0]{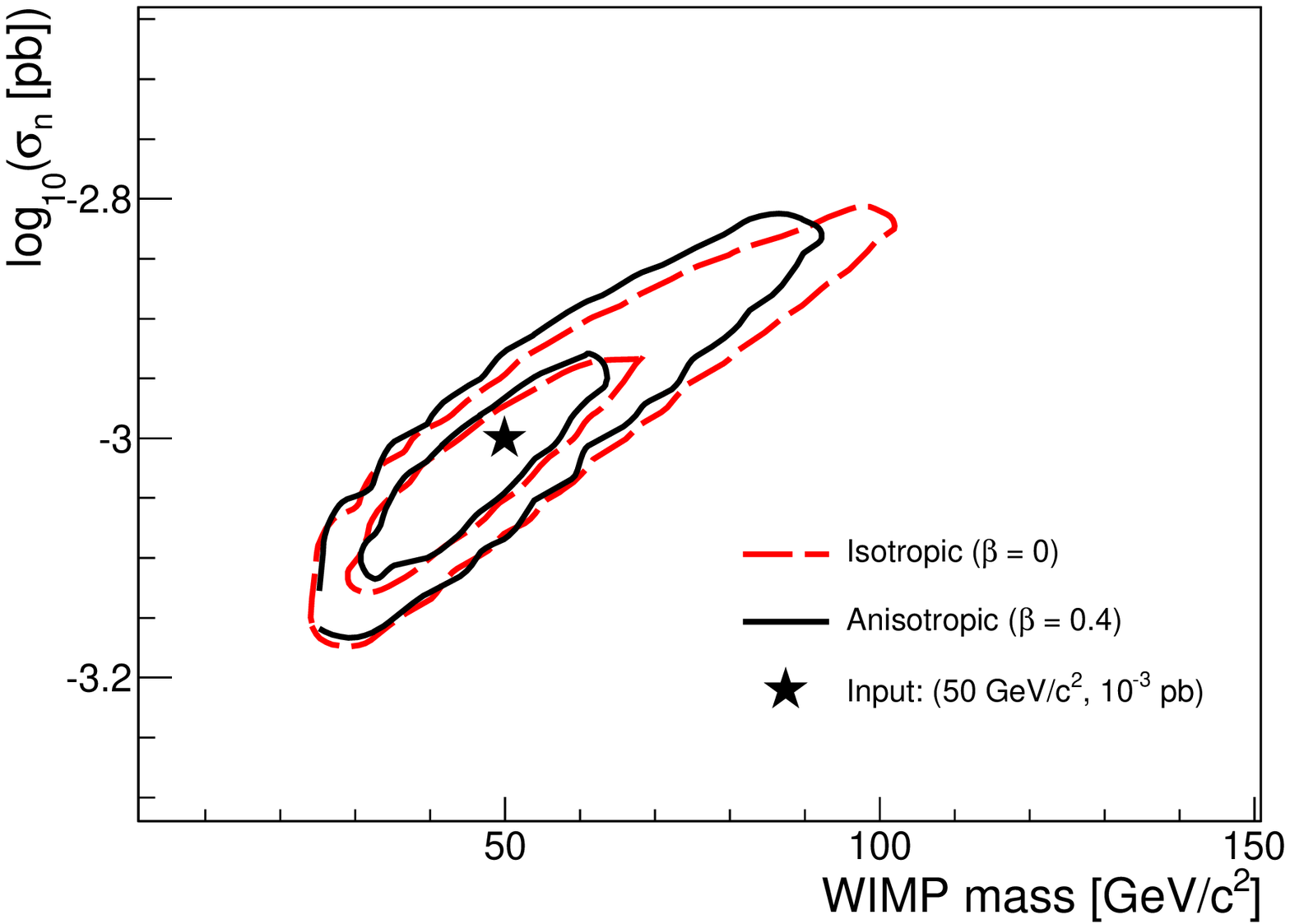}
\includegraphics[scale=0.39,angle=0]{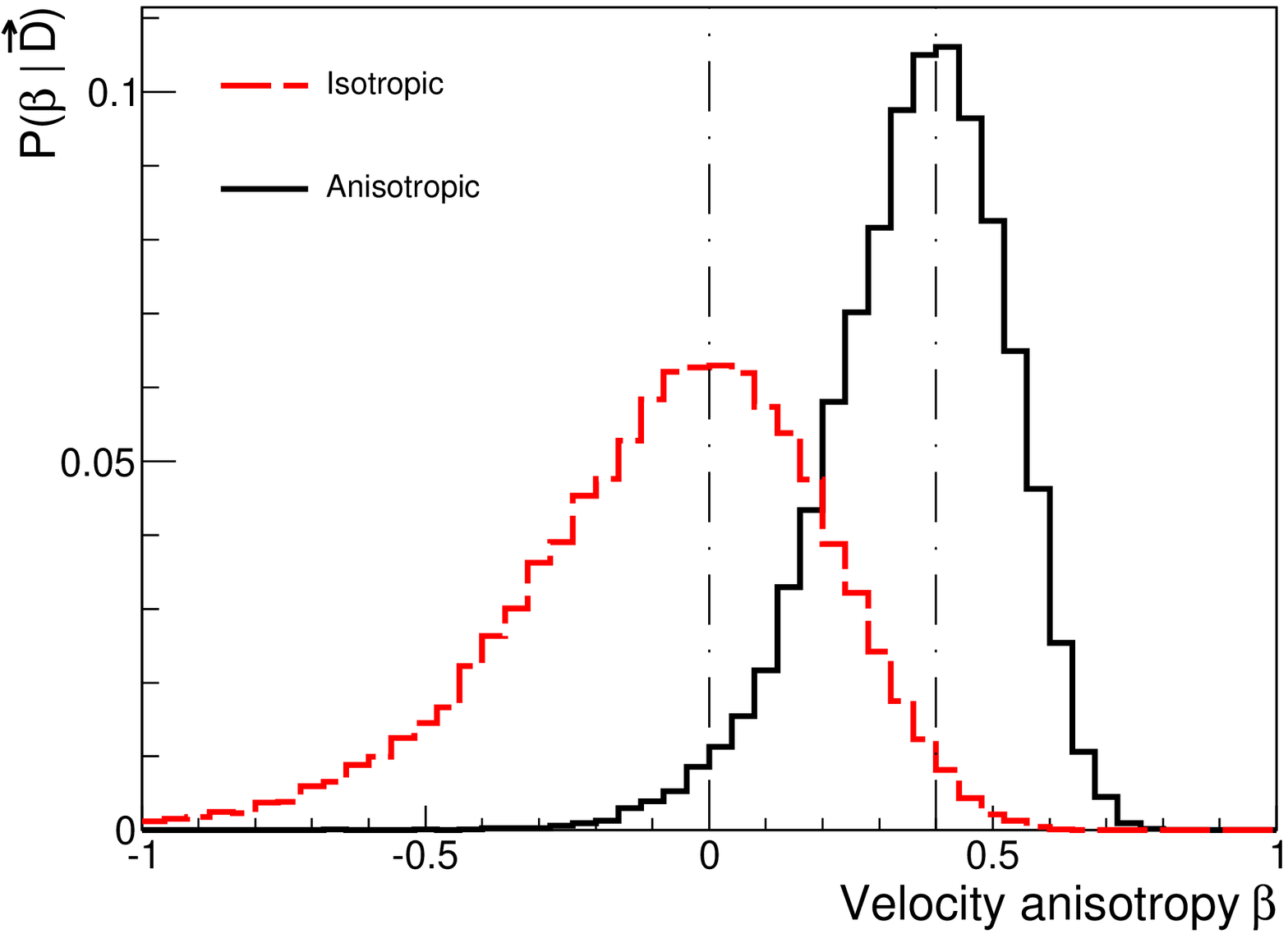}
\caption{Left panel : 68\% and 95\% contour level in the ($m_{\chi},\sigma_n$) plan, for a 50 $\rm GeV/c^2$ WIMP and for two input models : isotropic ($\beta=0$) and 
triaxial ($\beta=0.4$). 
Right panel : posterior PDF distribution of the $\beta$ parameter for the same models.}  
\label{fig:HaloT4}
\end{center}
\end{figure*}

From the previous section, we have conclude that this MCMC analysis tool is very efficient to constrain Dark Matter properties. However, in this section we will briefly
discuss the effect of some of the input parameters on the different constraints in order to estimate the performance of this analysis tool. 
We will first focus on the impact of the input WIMP mass. To do so, we have simulated
three different sets of directional data corresponding to an input WIMP mass of 
$m_{\chi} = 20, 50, 100 \ {\rm GeV/c^2}$ with a constant WIMP-nucleon cross-section  $\sigma_n = 10^{-3} \ {\rm pb}$ and the standard isotropic halo model. 
The results from the three MCMC runs are illustrated on figure \ref{fig:WIMPMass}. 
We present for the three WIMP masses, on the left panel,  the 68\% and 95\% CL contours in the 
($m_{\chi},\log_{10}(\sigma_n)$) plan and on the right panel, the posterior PDF $P(\beta|\vec{D})$ of 
the anisotropy velocity parameter $\beta$.\\
It can be deduced from figure~\ref{fig:WIMPMass} that the constraints strongly depend on the input WIMP mass, but in each case,
 the constraints are consistent with the input values. Then, this analysis has been shown to be working for any input WIMP mass although the constraints are stronger for
 light WIMPs. This is due to the fact that the signal characteristics, {\it i.e} the slope of  the 
energy distribution and the width of the angular distribution, evolve slowly with the WIMP mass once $m_{\chi} \geq 100$ GeV/c$^2$, as shown in \cite{billard.disco}.\\

In the following, we investigate the effect of an extremely triaxial halo model with $\beta=0.4$
on the estimation   of the Dark Matter parameters ($m_{\chi},\sigma_n,\beta$). The results from the MCMC run on a simulated 
dataset corresponding to a WIMP mass of 50 GeV/c$^2$ with the anisotropic halo model are presented on
figure \ref{fig:HaloT4}. As previously, on the left panel is presented the constraint at 68\% and 95\% on the 
($m_{\chi},\log_{10}(\sigma_n)$) plan while
 on the right panel is given the
deduced posterior PDF of the $\beta$ parameter. For convenience and comparison, 
the results from the benchmark input model (isothermal sphere with a 50 GeV/c$^2$ WIMP) are recalled.\\
From the left panel of figure \ref{fig:HaloT4}, we can conclude that the two halo models give similar constraints which are both consistent with the input values.
 In fact, and as foreseen, the fact that the velocity
dispersions are set as free parameters in the MCMC analysis allows to avoid induced bias due to wrong model assumption. 
From the right panel of figure \ref{fig:HaloT4} we can deduce that the $\beta$ parameter is well constrained: 
$\beta = 0.38^{+0.2}_{-0.1}$ and strongly in favor of an anisotropic Dark Matter halo.\\

As a conclusion of this study, it should be highlighted that the combination of   information from 
the angular and energy distributions leads to robust allowed regions in the ($m_{\chi},\log_{10}(\sigma_n)$) plan, since 
the halo model is also being constrained with the MCMC analysis from the same dataset of a single directional detection experiment. 
 Moreover, the velocity anisotropy parameter $\beta$, {\it i.e.} the three velocity dispersions,
  could be sufficiently constrained to discriminate between different halo  models with future directional detectors such as the one proposed by the MIMAC collaboration
  \cite{santos}.

\section{Track reconstruction with MIMAC}
\subsection{MIMAC prototype}

The MIMAC prototype is the elementary chamber of the future large matrix. It allows the possibility to show the ionization and track measurement performance needed to
 achieve the directional detection strategy.
The primary electron-ion pairs produced by a nuclear recoil in one chamber of the matrix are detected by drifting the electrons to the grid of a bulk micromegas \cite{bulk}
 and producing the avalanche in a very thin gap (128 or 256$\mu$m). 
 The electrons move towards the grid in the drift space and are projected on the anode thus allowing to get 
information on X and Y coordinates.
To access the X and Y dimensions with a 100 $\mu$m spatial resolution, a bulk micromegas  with a 4 by 4 cm$^2$ active area, segmented in pixels with a pitch of 350 $\mu$m
 is used as 2D readout.
 In order to reconstruct the third dimension Z of the recoil, a self-triggered electronics has been developed. It allows
  to perform the anode sampling at a frequency of 40 MHz.
This includes a dedicated 16 channels ASIC \cite{richer} associated to a DAQ \cite{bourrion}. 
The total recoil energy is deduced from the measured ionization quenching factor (IQF) \cite{santosQuenching}.\\

In order to fully exploit the data from the MIMAC detector and to constrain Dark Matter properties, the recoiling tracks have to be accurately reconstructed: direction, sense and
position in the detector volume. To achieve this goal, we first developed a track simulation software using SRIM \cite{srim} (simulated tracks), Magboltz \cite{Magb}
(electron drift velocity and dispersions) combined with the detector response simulation (MIMAC DAQ). This way, we can simulate any kind of recoiling track according to the following 
input parameters: X, Y, Z, $\theta$, $\phi$ and the sense ${S}$ (upward or downward). Each track is characterized by three types of topological observables: number of slices,
position of the center of gravity $X^{bary}$ and $Y^{bary}$ of the collected charge for each slice and the width $\Delta X^{bary}$ and $\Delta Y^{bary}$ along X and Y of
 each slice. Hence, the number of topological observables associated to any recoil track is given by:
\begin{equation}
N_{obs} = 1 + 4\times N_{slice}
\end{equation}
In this paper, we present a new MIMAC data analysis, based on \cite{billard.track}, in order to recover the track parameters: \{X, Y, Z, $\theta$, $\phi$, ${S}$\}. 
As a working example, we will consider in the following a simulated track of an hydrogen recoil of 100 keV at the position \{X = 0, Y = 0, Z = 5\} cm, in the direction
\{$\theta = 45^{\circ}$, $\phi = 0^{\circ}$\} and going downward. The analysis is based on the computation of the likelihood function
 $\mathscr{L}(X, Y, Z, \theta, \phi| {S})$ defined as,
 \begin{equation}
\mathscr{L}(X, Y, Z, \theta, \phi| {S}) = P(N_{slice})\prod_{n=1}^{N_{slice}}P(X_n^{bary})P(Y_n^{bary})P(\Delta X_n^{bary})P(\Delta Y_n^{bary})
 \end{equation}
where the different probability $P$ are estimated using a frequentist approach by simulating a large number of tracks for each step in the parameter space.

\setlength{\tabcolsep}{0.1cm}
\renewcommand{\arraystretch}{1.4}
\begin{table}
\caption{Comparison between the input and the output values of the track characteristics with the \{${S} =$ downward\} hypothesis for
 a 100 keV proton recoil in 50 mbar of pure isobutane. We quote mean value of the PDF distribution and (68 \% CL) error bars.}
{\lineup
\label{tab:modelinputTrack}
\begin{center}
\begin{tabular}{llllll}
\br
& \0\0\0\0\0X [cm] &  \0\0\0Y [cm] & \0\0\0Z [cm] & \0\0\0$\theta$ [$^{\circ}$] & \0\0\0\0\0$\phi$ [$^{\circ}$]  \\
\mr
Input &  \0\0\0\0\0\0\00  &  \0\0\0\0\00  & \0\0\0\0\05 & \0\0\0\045 & \0\0\0\0\0\00 \\
Output &  $-0.003\pm0.009$ & $0.006\pm0.012$  &  $5.068\pm0.18$ & $45.47\pm1.7$ & $-0.59\pm2.6$ \\
\br
\end{tabular}
\end{center}
}
\end{table}
\renewcommand{\arraystretch}{1.1}

In order to compute the likelihood function, we used a Markov Chain Monte Carlo algorithm and the full MCMC result following the \{${S} =
$ downward\} hypothesis is given in \cite{billard.track}. Deduced constraints on
 the five parameters and comparison with their input values are described in table~\ref{tab:modelinputTrack}. From this single track analysis
it can be deduced that the spatial uncertainty is of the order of 100 $\mu$m for X and Y and about 2 mm for the absolute height Z; and the angular uncertainty is about
3$^\circ$. However, even if systematic studies are needed to accurately estimate the different resolutions, results from table \ref{tab:modelinputTrack} are strongly in favor
of accurately reconstructed tracks.\\

So far, we did not discuss the possibility to discriminate between the two hypothesis: downward or upward, the so-called "head-tail effect". To do so with our track
reconstruction method, we compute the likelihood function according to the second hypothesis \{${S} =$ upward\}. 
Then, a direct comparison of the maximum Log-Likelihood values
of each hypothesis (-3.8 for \{${S} =$ downward\} and -5.1 for \{${S} =$ upward\}) allows to conclude in favor of a downwarding track consistent with the input
simulated track. Hence, this method seems to be able to achieve the sense recognition of the recoiling tracks, but here again, systematic studies must be done to estimate its
efficiency \cite{billard.track}.

\section{Conclusion}

In this paper, we have shown that the directional detector proposed by the MIMAC collaboration should be sensitive to an important region of interest of the Dark Matter
parameter space ($m_{\chi}$, $\sigma_n$) motivated by constrained minimal supersymmetric models. In the case of a strong WIMP positive detection with $\sigma_n \sim 10^{-3}$
pb, an identification of Dark Matter could be achieve in a model independent way using the MCMC analysis method presented in section 2. We have shown that constraining 
the velocity dispersions of the galactic Dark Matter halo allows to obviate bias in the determination of the WIMP properties, 
due to wrong halo model assumptions. Moreover, we have shown the possibility
to discriminate between various halo model characterized by their velocity anisotropy parameter. However, such interesting results requires accurate track reconstruction:
 direction, position and sense. To reach this goal, we developed a new track reconstruction method based on a
likelihood analysis which has shown very promising results, {\it i.e} high resolutions. Although, systematic studies must be done to accurately estimate the differents
resolutions and the sense recognition efficiency.

\end{document}